# Disentangling the frequency content in optoacoustics

**Antonia Longo[a,b,c+], Dominik Jüstel[a,b,d+], Vasilis Ntziachristos[a,b*]**

[a] Chair of Biological Imaging at the Central Institute for Translational Cancer Research (TranslaTUM), School of Medicine, Technical University of Munich, Ismaninger Str. 22, 81675, Munich, Germany

[b] Institute of Biological and Medical Imaging, Helmholtz Zentrum München, Ingolstädter Landstr. 1, D-85764, Neuherberg, Germany

[c] iThera Medical GmbH, Zielstattstrasse, 13, 81379, Munich, Germany

[d] Institute of Computational Biology, Helmholtz Zentrum München, Ingolstädter Landstr. 1, D-85764, Neuherberg, Germany

[+] These two authors contributed equally.

[*] Corresponding author: Vasilis Ntziachristos, bioimaging.translatum@tum.de

Contact info: Antonia Longo, antonia.longo1@gmail.com; Dominik Jüstel, dominik.juestel@helmholtz-muenchen.de; Vasilis Ntziachristos, bioimaging.translatum@tum.de.

## Abstract

Signals acquired by optoacoustic tomography systems have broadband frequency content that encodes information about structures on different physical scales. Concurrent processing and rendering of such broadband signals may result in images with poor contrast and fidelity due to a bias towards low frequency contributions from larger structures. This problem cannot be addressed by filtering different frequency bands and reconstructing them individually, as this procedure leads to artefacts due to its incompatibility with the entangled frequency content of signals generated by structures of different sizes. Here we introduce frequency-band model-based (fbMB) reconstruction to separate frequency-band-specific optoacoustic image components during image formation, thereby enabling structures of all sizes to be rendered with high fidelity. In order to disentangle the overlapping frequency content of image components, fbMB uses soft priors to achieve an optimal trade-off between localization of the components in frequency bands and their structural integrity. We demonstrate that fbMB produces optoacoustic images with improved contrast and fidelity, which reveal anatomical structures in *in vivo* images of mice in unprecedented detail. These enhancements further improve the accuracy of spectral unmixing in small vasculature. By offering a precise treatment of the frequency components of optoacoustic signals, fbMB improves the quality, accuracy, and quantification of optoacoustic images and provides a method of choice for optoacoustic reconstructions.

# Introduction

In contrast to ultrasonography, optoacoustic (OA) imaging systems typically generate acoustic signals with a rich frequency content due to the ultra-short illumination pulses. The broad frequency content carries information about structures on different physical scales (i.e., size). Therefore, advanced OA systems use broadband acoustic detectors to ensure that no structural information is lost. Despite the rich information that is contained in the frequency spectrum, typical OA reconstructions do not consider the particular dependencies of image features on frequency. For example, the power of the OA signals scale with the size of the source, resulting in a frequency dependent signal-to-noise ratio (SNR) and images where small structures may be obscured by the high SNR of low frequency contributions (see Supplementary Information and Supplementary Fig. 1c,d). Processing and rendering all frequencies together therefore bias the visual perception of an image towards larger structures. Furthermore, the sensitivity fields of broadband ultrasound detectors employed in state-of-the-art OA systems are also frequency dependent (see Supplementary Information and Supplementary Fig. 1a) [1]; lower frequencies sample larger volumes, which, in the case of linear ultrasound arrays, may appear as out-of-plane signals in the reconstructed image (see Supplementary Information and Supplementary Fig. 1.e,f) and further obscure features formed by higher frequencies.

Current OA image reconstruction techniques do not accurately extract all the information contained in broadband OA signals. Common filtered back-projection methods [2] act as high-pass filters on the image, while delay-and-sum algorithms [3] and model-based algorithms [4, 5] treat all frequencies equally, thereby penalizing the high frequencies and leading to less accurate rendering of smaller structures. Both methods result in an oxymoron; they afford visually pleasing images by penalizing either low or high frequency content and ultimately reducing image fidelity.

Separate reconstruction of different frequency bands has been considered in Ultra-broadband Raster-Scan Optoacoustic Mesoscopy (RSOM) [6, 7] and in OA microscopy [8]. However, simple separation of the frequency bands using signal processing filters neglects the entanglement of the broadband frequency content of signals that are generated by structures of different sizes, which may result in artefacts (e.g., ring artefacts) that can be further amplified by limited view detection [9, 10] or insufficient spatial sampling [10, 11]. Moreover, model-based reconstruction techniques for OA tomography systems have yet to account for the abovementioned dependence of image content on the recorded acoustic frequencies. Out-of-plane signals have been separated from in-plane signals by axially displacing the transducer array and by de-correlating images acquired at different out-of-plane positions with the in-plane image [12, 13]; however, such approaches are impractical since they require the displacement of the transducer and do not consider the frequency-dependence of the detector's sensitivity.

In this work, we propose a frequency-band separated model-based (fbMB) reconstruction framework for OA tomography that untangles the overlapping frequency content of image components to improve resolution and overall image fidelity. The fbMB method reconstructs multiple images such that each is a partial model-based reconstruction of a portion of the signal and that these images sum up to a reconstruction of the full OA signal data. fbMB imposes soft priors on the frequency content of the signals underlying the different image components to allow integration of frequencies from other bands if required by the model, thereby achieving an optimal trade-off between structural fidelity of these components and localization of the corresponding signals in separate frequency bands. Lastly, fbMB imposes non-negativity constraints on all image components to ensure that artefacts in one component cannot be compensated by negative contributions in others. Furthermore, fbMB can be extended to any image reconstruction problem that is based on a linear model of physical phenomena and the computational complexity only increases linearly with the number of components compared to a standard model-based algorithm. We demonstrate on both a synthetic phantom and real OA data that, with a suitable choice of filters depending on the biological application, the new method can efficiently separate structures of different physical scales, as well as out-of-plane from in-plane signals. Using a dataset of 533 *in vivo* multispectral OA tomography (MSOT) images of different anatomical sections of mice, we show that the new method generates OA images with lower reconstruction residuals and higher structural similarity index values compared with simple filter techniques, and improved image contrast and resolution compared with a standard model-based algorithm. The fbMB method implements a unique multi-frequency-band contrast in MSOT to complement its volumetric imaging, real-time acquisition, and multispectral contrast capabilities. By rendering the rich, high-fidelity OA content from biological structures on all scales, fbMB could expand the applications of MSOT in both preclinical and clinical medicine.

# Results

The proposed fbMB method separates OA signal contributions from different frequency bands by integrating the filtering process with soft priors into the reconstruction model, and by applying non-negativity constraints on all components. In the resulting multiscale images, anatomical structures of different sizes are more accurately represented. A schematic of two-band model-based reconstruction (2bMB) is provided in Fig. 1a. The 2bMB reconstruction is applied to the OA signal acquired by the system and generates OA images of specific frequency band components: a low frequency image (Band 1, Image domain in Fig. 1a) containing mainly signals from larger structures and a high frequency image (Band 2, Image domain in Fig. 1a) corresponding to smaller structures. OA images with multiscale contrast are generated by color-coding and blending the different components (2bMB, Image domain in Fig. 1a). Notably, the soft priors on the frequency content of the components allow each component to include frequencies from other band if the model requires it (2bMB, Signal domain in Fig. 1a).

In the following, we report our findings in phantoms and in *ex vivo* and *in vivo* mouse data. This data is supported with figures of merit, such as the reconstruction residuals, the structural similarity index measure (SSIM), the image entropy and the Perception-based Image Quality Evaluator (PIQE).

Section 1 details the effectiveness of the proposed method to separate different frequency band components in OA imaging. Sections 2 and 3 describe the validation of the method against standard signal filtering and simple model-based reconstruction. Section 4 reports spectral unmixing results to demonstrate multiscale quantification of specific chromophores.

## Disentanglement of frequency content by fbMB

The capability of the algorithm to generate images with band-specific contrast is first demonstrated on a controlled synthetic phantom with well-defined features of different sizes (Fig. 1b). The phantom features a network of vessel-like structures with diameters ranging from 1.3 mm to 240 μm, which are embedded in a circular absorbing background that is 1.7 cm in diameter and absorbs 10 times less than the vessel-like structures (see Methods section). We tested two-band decomposition by applying a two-band model-based (2bMB) algorithm. For 2bMB, two band-pass filters with frequency bands 0-1.23 MHz and 1.23-15 MHz were chosen for the soft priors, with 175% relative bandwidth (see Methods section).

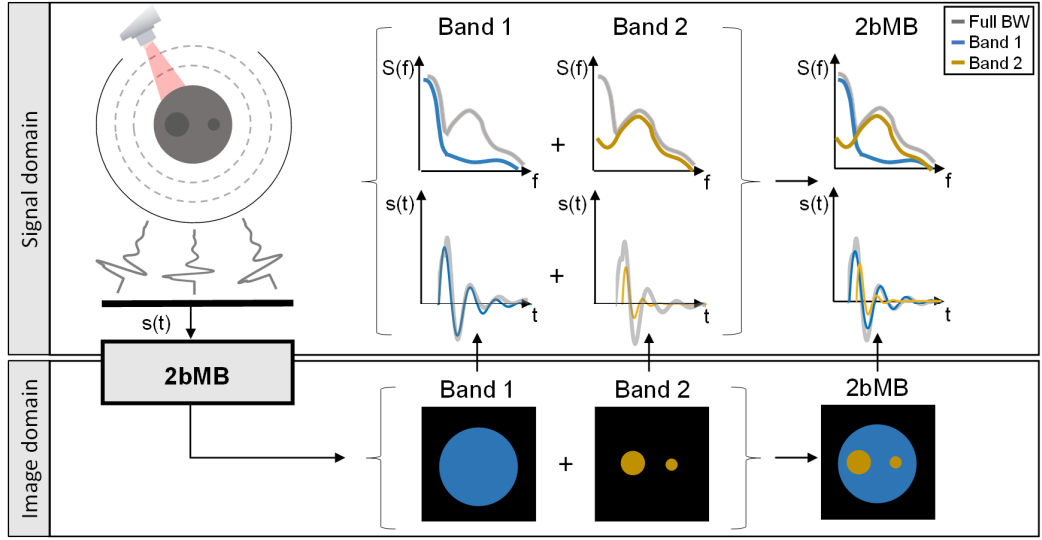

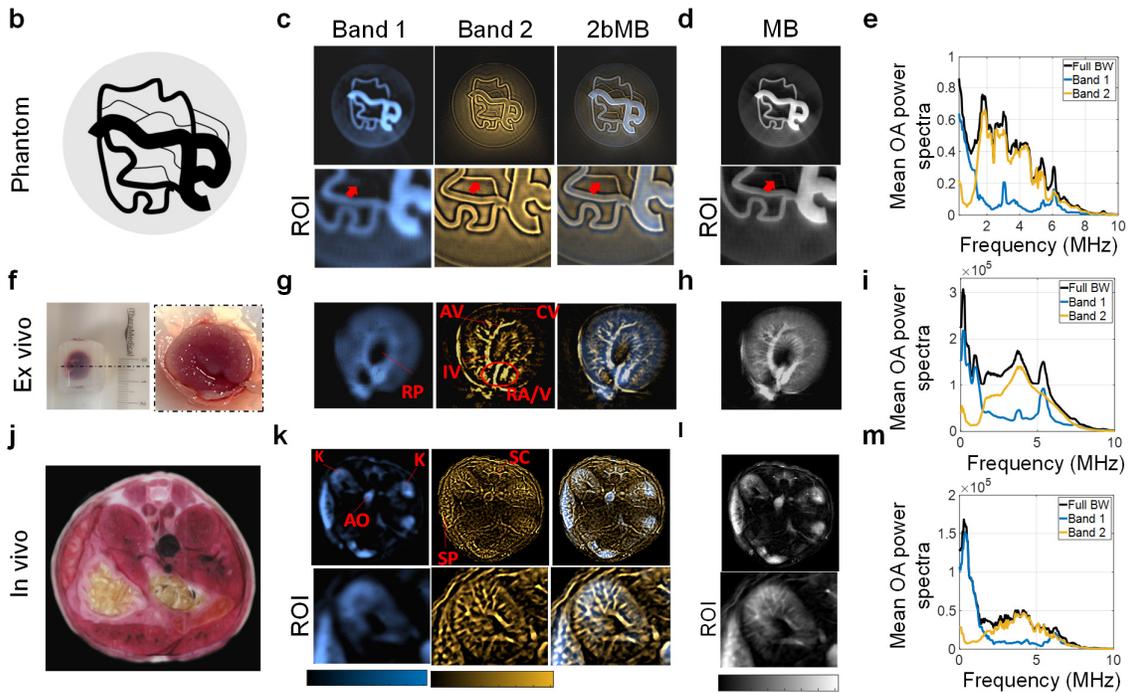

**Figure 1. Frequency content disentanglement in OA imaging and frequency-band model based (fbMB) reconstruction. a)** Schematic of two-band model-based reconstruction algorithm (2bMB). **b)** A synthetic phantom with well-defined features of varying size. Reconstructed OA images of the phantom for **c)** two-band model-based and **d)** standard model-based reconstruction (MB). The 2bMB image was obtained by color-coding and blending the different frequency band components. The red arrows in the magnified ROIs indicate a small structure enhanced with the new method. **e)** Mean power spectra of the OA signals corresponding to the two different frequency band components of the phantom showing frequency overlaps between the components due to soft priors in the reconstruction. **f)** Ex vivo mouse kidney embedded in agar and magnification of the selected anatomical cross section in the middle segment. OA images of the kidney cross section reconstructed using **g)** two-band model-based and **h)** standard model-based reconstruction. **i)** Mean power spectra of the OA signals corresponding to the two frequency band components. Visible anatomical details in the vascular network of the excised kidney for 3bMB: Main renal vessel (RA/V), interlobal (IV), arcuate (AV) and cortical (CV) vessels. **j)** Anatomical image of a mouse kidneys section including spleen and liver lobes. In vivo OA images of the mouse kidney section reconstructed by using **k)** two-band model-based and **l)** standard model- based reconstruction.

Kidney (K), aorta (AO), spleen (SP) and spinal cord (SC). OA frequency content for the corresponding frequency band components for m) two-band decomposition.

Fig. 1c shows images of the phantom that were reconstructed by the two-band model-based (2bMB) algorithm. The new algorithm successfully decomposes images into components on different physical scales. Low frequency images (Fig. 1c, Band 1) contain mainly signals from larger structures ranging from 1.6 cm to 600 µm in diameter, while high frequency images (Fig. 1c, Band 2) correspond to small structures that are less than 600 µm across. Color-coding and blending the frequency band components generates OA images with multiscale contrast (Fig. 1c, 2bMB). Compared with standard model-based reconstruction (MB in Fig. 1d), 2bMB affords a more accurate image and improved visualization of small features (red arrows in the ROIs).

Fig. 1e reports the mean power spectra of the OA signals generated by the structures on different scales for the 2bMB algorithm, which demonstrate that the frequency contents of the different reconstructed components overlap significantly. This finding confirms that the new method achieves a more realistic disentanglement of different-sized structures compared with simple frequency filtering, which cannot model such overlapping frequency content. The separation of this overlapping content is enabled by the soft priors on the filtering process in the reconstruction, which allows the components to integrate frequencies from other bands if the model requires it (see Methods section).

Next, the algorithm was applied to OA images of an *ex vivo* mouse kidney (Fig. 1f) to demonstrate the ability of frequency band contrast to reveal structures and vessels of different sizes in tissue. Fig. 1g shows an OA image after decomposition into two bands. As with the phantom, the algorithm efficiently separated the different anatomical components in the kidney. Low frequency components contain primarily OA signals from the organs and large vessels. High frequency components have better in-plane focus and show small vessel networks with high accuracy. Furthermore, as shown in Fig. 1i the OA spectra of the different bands overlap in the frequency domain, again confirming the effect of the soft priors on the reconstructed images, which enable each component to integrate frequencies from other bands to assure structural and consistency with the underlying physics. Blending the two bands (2bMB in Fig. 1g) afforded OA images in which fine details, such as small vasculature inside bulk tissue, were rendered with significantly higher contrast than possible with standard reconstruction (Fig. 1h).

The frequency disentanglement approach was then applied to reconstruct an image of a kidney section in a live mouse (Fig. 1j) to demonstrate image quality improvement and structural enhancement *in vivo*. Fig. 1k shows the different frequency band components of the mouse kidney section, obtained by applying 2bMB. The low-frequency bands contain signals from organs, such as the kidneys (K) and spleen (SP), or large structures, such as the abdominal aorta (AO), and the spinal cord (SC). The high-frequency components contain signals from small vessels, such as the microvasculature in the kidney, the splenic vessel (SPV), and the hepatic vessels (HEV). As expected, the OA signals from these structures again overlap in the frequency domain (Fig. 1m), supporting the physical fidelity of the reconstructed *in vivo* images. Finally, the OA images obtained by blending the different frequency band components (2bMB in Fig. 1k) visualize with high contrast the entire OA information content, which is not possible with the standard model-based reconstruction (MB in Fig. 1l).

Finally, we tested three-band model-based (3bMB) reconstruction in order to evaluate the effect of a finer frequency band decomposition on the image contrast. For 3bMB, three band-pass filters with frequency bands 0-500kHz, 500kHz-1.23MHz and 1.23MHz-13.3MHz were selected for the soft priors, with a relative bandwidth of 136% (see Methods). Similar to 2bMB, 3bMB successfully separates the different OA frequency band components in the images. Low frequency images contain signal from bulky absorbing tissue and large structures, while the high frequency images contain signals from smaller structures (Fig. 2a,c,e). Furthermore, as shown in Fig. 2b,d,f, the individual bands overlap consistently in the frequency domain due to the soft priors in the reconstruction.

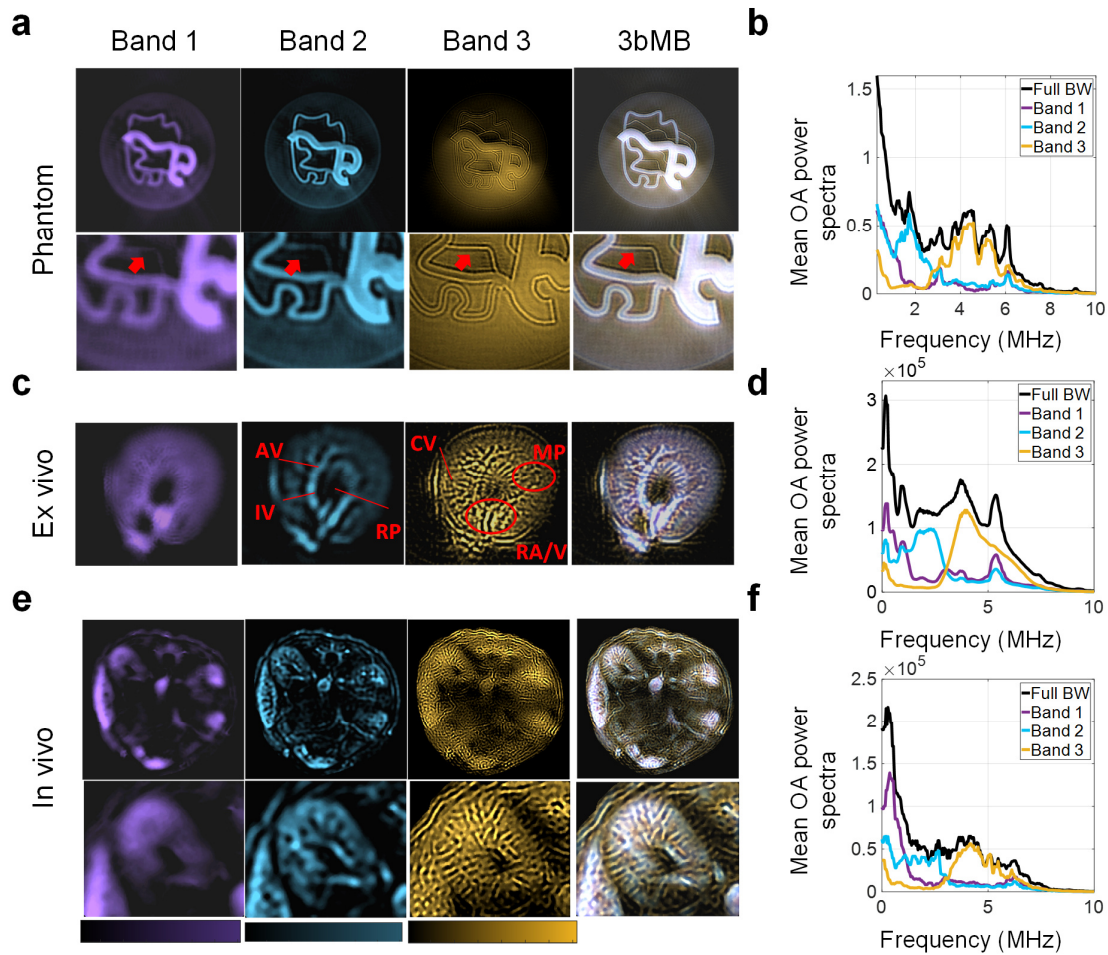

**Figure 2. 3bMB reconstruction**. a) Three-band decomposition of OA images of the synthetic phantom from Fig. 1b. The 3bMB images were obtained by color-coding and blending the three components. The red arrows in the magnified ROIs indicate a small structure enhanced with the new method. b) Mean power spectra of the OA signals of the phantom for three different frequency band components with frequency overlaps between the components due to soft priors in the reconstruction. c) 3bMB reconstructions of the excised kidney cross section shown in Fig. 1f. d) Mean power spectra of the OA signals corresponding to the three frequency band components. Visible anatomical details in the vasculature network of the kidney for 3bMB: Main renal vessel (RA/V), interlobal (IV), arcuate (AV) and cortical (CV) vessels. Renal cortex (RC), medullary pyramids (MP). e) *In vivo* 3bMB reconstructions of the mouse kidney section of Fig. 1j. f) OA frequency content of the three frequency band components.

By increasing the number of bands from two to three, we observed that the high frequency structures are further decoupled from the middle and low frequencies, thus demonstrating that by finer decomposition of the signal, the image resolution further increases, and the visual perception of small structures improves. For example, the three-band reconstruction of the ex-vivo kidney section in 3bMB in Fig. 2c reveals further anatomical details of the vascular network of the kidney with high accuracy. A notable example is the main renal vessel (RA/V) entering the kidney and progressively branching into the interlobal (IV), arcuate (AV) and cortical (CV) vessels, which cannot be identified in the 2bMB image (2bMB in Fig. 1g) or the standard model-based reconstructed image (MB in Fig. 1h). 3bMB thus allows detailed visualization of the entire blood supply in the kidney.

## Comparison of 2bMB against simple filtering techniques

Compared to standard filtering methods, the proposed algorithm affords images with significantly higher physical and structural accuracy. If an image is decomposed into a superposition, each component should comply with the underlying physics. We can quantify the physical accuracy by comparing the estimated residual errors, i.e., the difference between the recorded OA signals and the model of the imaging system applied to the reconstructed image, which the sum of the frequency band components. In addition, the superposition of the different components should produce an image without introducing artefacts (e.g., ripples at edges). Structural fidelity between two images can be quantified using the structural similarity index measure (SSIM), which was used herein to compare the output of the proposed algorithm to that of standard model-based reconstruction. In this way, we validated the physical and structural integrity of the fbMB reconstruction against standard signal filtering with Butterworth filters and Wavelet filters (Fig. 3) on a dataset of 533 OA images from different anatomical sections.

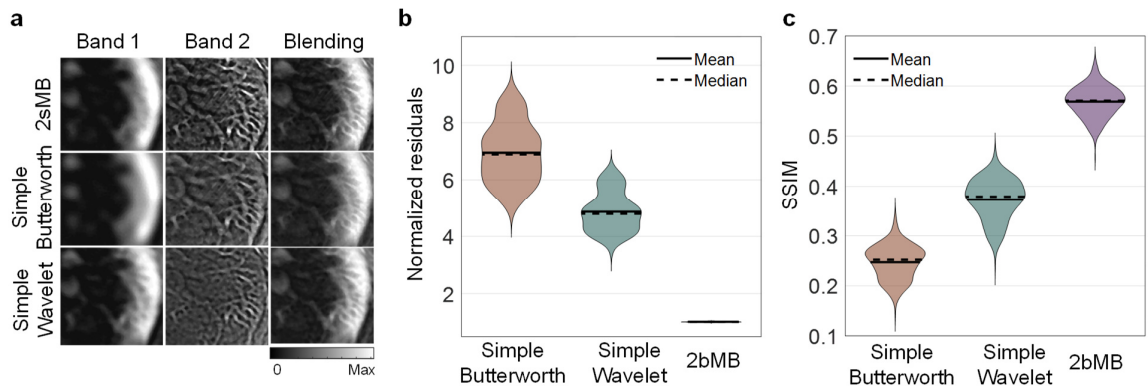

**Figure 3. Comparison of 2bMB against simple filtering techniques.** a) Reconstruction of an in vivo mouse liver section using 2bMB (top), simple Butterworth (middle), and SWT (bottom) filtering techniques. b) Normalized reconstruction residuals and c) structural similarity index measure (SSIM) of 2bMB and simple filtering techniques, compared to standard model-based reconstruction.

Fig. 3a displays exemplary multiscale OA images from a mouse liver section obtained with 2bMB reconstruction and with model-based reconstruction after the application of Butterworth and Wavelet bandpass filters. All three methods separate different frequency bands of OA contrast in the liver section, highlighting bulky tissue (band 1) or vasculature (band 2). However, as shown in Fig. 3b, 2bMB generates OA images with lower normalized reconstruction residuals (mean of 1.3) compared to Wavelet (mean of 4.7) and Butterworth (mean value of 6.8). The residuals are normalized relative to those obtained with standard model-based reconstruction (see Methods section). Therefore, the new proposed decomposition approach has only a small regularizing effect when compared to standard model-based reconstruction, while the other methods deviate from the data significantly. Hence, the new method assures a solution with the highest physical accuracy, wherein the sum of the different OA components most accurately reflects the original OA signals.

In addition, as shown in Fig. 3c, the new algorithm generates images with the highest structural similarity index measure SSIM of 0.6 when compared with the reference reconstruction, while the filtering techniques achieve values of 0.39 (Wavelet) and 0.21 (Butterworth), respectively. Therefore, integrating the filtering process with soft priors into the inversion assures that no artefacts are generated by the decomposition and that the integrity of the structure being imaged is better preserved. The still low value of structural similarity of 0.6 for the 2bMB reconstruction is due to the regularizing effect of the frequency band decomposition.

**Comparison 2bMB against standard model-based reconstruction**

Fig.4 demonstrates the effectiveness of 2bMB to separate anatomical structures on different scales in *in vivo* whole-body images of a mouse. The improved contrast of 2bMB image (obtained by color-coding and blending images at different frequency bands) was quantified against a standard model based reconstructed image (MB) using two figures of merit: the image entropy and the Perception-based Image Quality Evaluator (PIQE).

Fig. 4a shows the maximum intensity projection (MIP) images of the mouse in the YZ and XZ planes for two-bands decomposition (Band 1 and Band 2) reconstructed with 2bMB algorithm. As expected, the low frequency image contains OA signals from large anatomical structures, such as organs, muscles, and large vessels, while the high frequency image shows the vasculature network inside of the mouse. Several anatomical structures can be identified in the low-frequency band image (blue), including the carotid artery and the jugular vein in the mouse neck, the aorta and the vena cava running parallel in the mouse abdomen and branching into the iliac artery/vein in the lower abdomen, the trapezius muscle in the upper part of the spine, the femoral muscle in the lower extremities of the mouse, the kidneys, and the liver lobes in the abdomen. Visible in the high-frequency band image (yellow) are the dorsal intercostal veins and arteries between the ribs, which provide blood supply in the thoracic wall, and the superficial epigastric artery and vein, both cranial and inferior. Fig. 4b shows the image quality improvement of 2bMB in both 3D and 2D mouse sections, compared to standard model-based reconstruction (MB). The 2bMB images, obtained by color-coding and blending the different bands (Fig. 4a), display more anatomical details and broadband OA contrast than OA images from standard MB reconstruction, which appear blurred and lack contrast for small structures. For example, the dorsal artery and vein are clearly visible in the 2bMB image but hidden in the MB images due to the proximity of liver and lungs. Fig. 4c,d report the values of entropy and PIQE estimated for 533 *in vivo* OA images from different mouse anatomical sections, which quantify the image quality improvement of 2bMB compared to standard MB reconstruction in a large dataset. The OA images reconstructed with the proposed method exhibit increases in entropy of 50%, higher visual contrast, and a reduction in the PIQE index of 19%, implying an overall improved visual image quality.

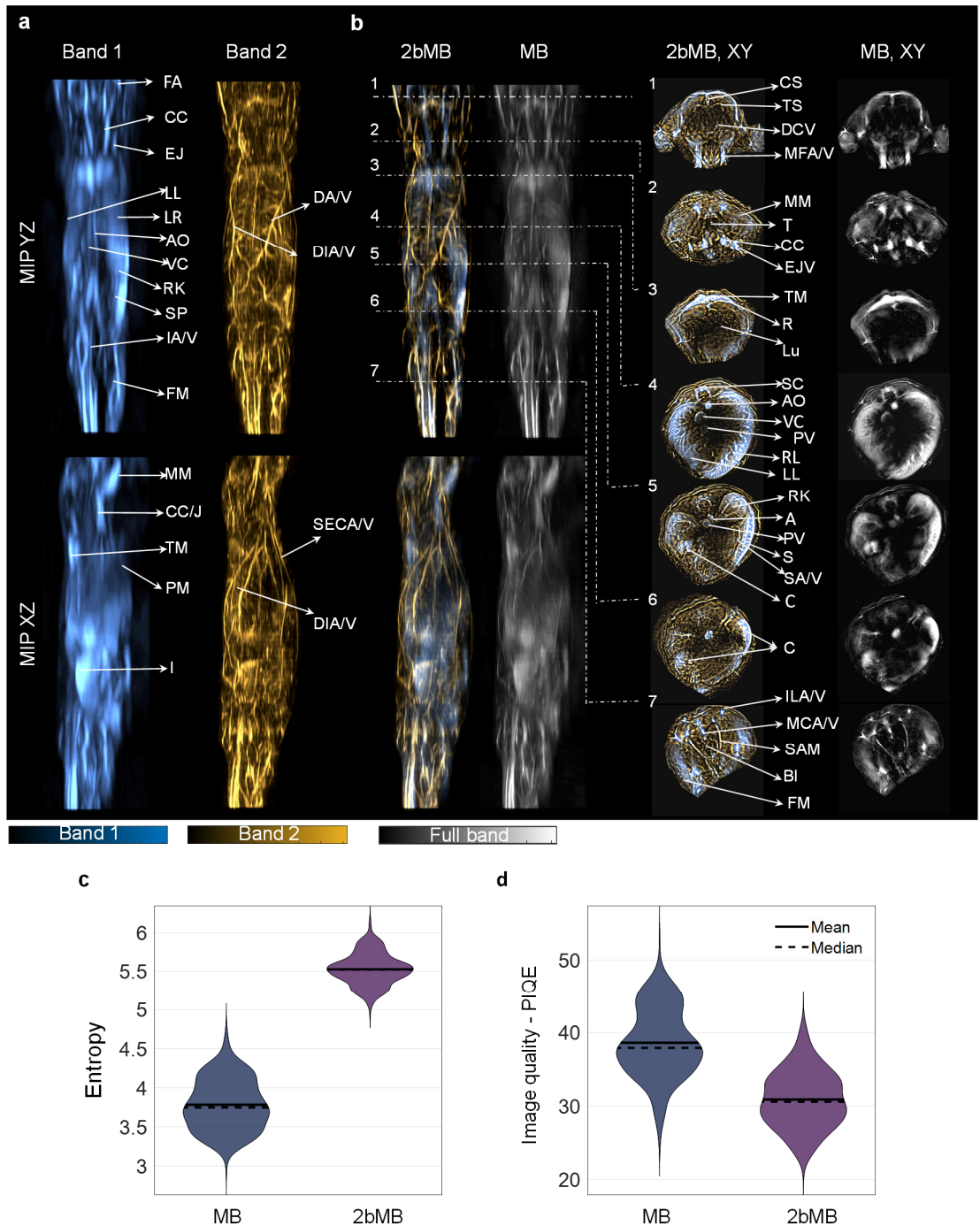

**Figure 4. Comparison of 2bMB against standard MB.** a) Maximum Intensity Projection (MIP) images of volumetric OA mouse data in the YZ and XZ planes for the two frequency band components obtained with 2bMB reconstruction. b) Comparisons of 2bMB reconstructions to standard model-based (MB) reconstructions for both volumetric data (left) and 2D sections (right) of the mouse. Image quality indicators c) entropy and d) Perception-based Image Quality Evaluator (PIQE). 1 brain section; 2 neck section; 3 lungs section; 4 liver section; 5 kidneys section; 6 colon section; 7 lower abdomen. Facial artery (FA); Common carotid (CC); External jugular (EJ); Liver left (LL); Liver right (LR); Aorta (AO); Vena cava (VC); Right kidney (RK); Spleen (SP); Iliac artery and vein (IA/V); Femoris muscle (FM); Masseter muscle (MM); Common carotid/ jugular (CC/J); Trapezius muscle (TM); Pectoral muscle (PM); Intestine I; Dorsal artery and vein (DA/V) Dorsal intercostal artery and vein (DIA/V);

Superficial epigastric cranial artery and vein (SECA/V); Confluence sinus (CS); Transverse sinus (TS); Deep cerebral vessel (DCV); Mandibula-facial arteries and veins (MFA/V); Trachea (T); Rib (R); Lungs (Lu); Spinal cord (SC); Splenic artery and vein (SA/V); Colon (C); Middle caudal artery and vein (MCA/V); Straight abdominal muscle (SAM); Bladder (Bl).

## Spectral unmixing

2bMB also improves image quantification via spectral unmixing in OA imaging by utilizing the regularization effect of the decomposition to prevent negative values of low frequency signals from compromising the signal quantification in small vasculature. Fig. 5a displays the absorption spectra of oxygenated hemoglobin (HbO$_2$) and deoxygenated hemoglobin (Hb) in the range 800 – 980 nm. Fig. 5b shows the Hb and HbO$_2$ spectral linear unmixing results for mouse cross sections of liver and kidneys, obtained from both 2bMB and standard MB reconstruction. Notably, the proposed method enables Hb and HbO$_2$ quantification in small vessels, which are otherwise obscured by low frequency signals. The magnification of a region-of-interest (ROI) in the right lobe of the liver clearly shows the small vasculature that supply blood to the liver; the magnification of an ROI in the left kidney demonstrates Hb and HbO$_2$ quantification in the splenic vessel, as well as in the small vasculature in the kidney cortex.

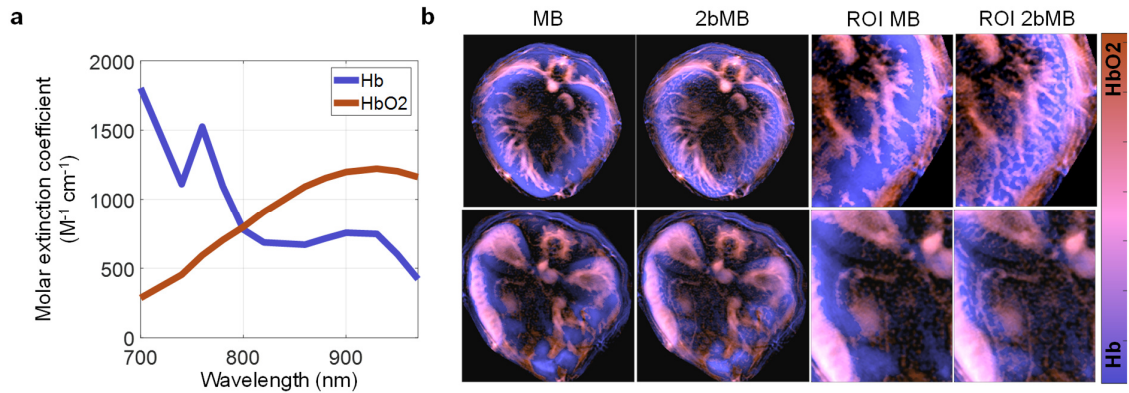

**Figure 5. Spectral linear unmixing of in vivo mouse liver and kidneys sections.** a) Absorption spectra of oxygenated hemoglobin (HbO$_2$) and deoxygenated hemoglobin (Hb) in the range 700 – 980 nm. b) Linear spectral unmixing of Hb and HbO$_2$ in mouse liver and kidneys sections obtained from OA images reconstructed with MB and 2bMB. Magnification of two ROIs in the left lobe of the liver and in the kidney-spleen region.

# Discussion

OA imaging yields rich information content due to the detection of broadband acoustic signals. However, it is challenging to faithfully render objects in different size regimes with a single image, which compromises the accuracy of subsequent physical and spectral analyses. To overcome this challenge, we developed fbMB, a method which employs soft priors on the frequency bands to generate OA images with frequency-band-specific contrast; low-frequency images contain bulk tissue, organs, and out-of-plane signals, while high-frequency images contain small vasculature and anatomical details. The method is a practical solution to the longstanding problem to reliably separate OA contrast from different frequency bands and thereby counteract the bias towards low frequencies in OA imaging.

Contrary to standard back-projection and model-based reconstructions, the new algorithm allows, for the first time, access to the full structural content of broadband OA images, which is of great value for preclinical and clinical applications. In combination with the three spatial dimensions (3D), the time dimension (4D), and the optical wavelength dimension (5D), fbMB adds a sixth "ultrasound frequency dimension" (6D) to OA imaging, thus enabling analysis of images over multiple scales [14]. The acquisition and visualization of six-dimensional OA data enables the full range of OA contrast mechanisms, i.e., real time volumetric imaging at multiple wavelengths of structures with a wide range of sizes, offering unique performance among current biomedical imaging modalities.

The ability of fbMB to separate different frequency bands without generating artefacts was quantitatively evaluated on a dataset of 533 *in vivo* OA images from different mouse anatomical sections. By imposing soft priors on the filters and allowing frequency bands overlap, the new method provides a unique solution to the decomposition problem, which is less prone to reconstruction errors and artefacts than rigid filtering. Moreover, fbMB generates OA images with richer informational content and better visual and structural perception compared to standard model-based reconstruction. The resulting OA images are more detailed and resolve anatomical structures with high biological accuracy, increasing the biomedical value of OA imaging systems.

Spectral unmixing applied to fbMB reconstructions enables quantification of absorbers at resolutions ranging from millimeters to micrometers. In general, spectral unmixing algorithms are applied to OA images reconstructed with standard model-based algorithms due to their higher accuracy compared to back-projection algorithms. However, a common problem for OA image quantification is that negative values in low frequencies of the signal originating from larger structures or artefacts [8] compromise the quantification of small-scale features. Here instead, the multiscale representation of different chromophores through fbMB prevents negative values from low frequencies to interfere with the signal from smaller structures and enables label-free quantification of the vasculature down to the capillary level. This result may expand applications of OA image quantification in, for example, cardiovascular imaging by providing anatomical information from macro- to micro- vasculatures, in cancer imaging by detecting angiogenesis and tissue perfusion, and in neuroimaging by mapping the hemodynamics in the brain vasculatures with high resolution.

The fbMB method is flexible and robust with respect to the choice of the filters and allows a broad range of applications and extensions. For instance, beyond multiscale contrast and numerical focusing (i.e., separating the out-of-plane from the in-plane signal), one could include the spectral dimension into the reconstruction, which could enable simultaneous unmixing for multiscale quantification of specific chromophores in the same schema. A variation of the introduced method could also help to decouple absorption and light fluence, which exhibits a global low spatial frequency component [15], thus potentially improving OA signal quantification in small vasculatures and in bulk tissue.

In addition, a dedicated filter design in the soft priors could improve the current results for special applications. Indeed, each of these frequency bands has biological and clinical potential, ranging from assessing mechanisms in organs and bulky tissues at the macroscopic scale to detecting physiological processes at microscopic levels (e.g., in the microvasculature).

Notably, the method is also applicable to other image reconstruction problems that suffer from the same bias towards low frequencies. Furthermore, the method can be easily rescaled and extended to OA microscopy and mesoscopy.

Recently, applications of machine learning in imaging have afforded real time implementations of previously computationally expensive reconstruction methods. The acceleration of fbMB through deep learning, would allow real time feedback of multiple frequency band components to the system user in clinical applications. This would enhance the clinical usage of MSOT for screening and diagnosis.

The proposed method enables a unique frequency-band contrast for OA imaging encoding the entire OA content in a single image. The improved image quality and visual perception of anatomical structures will advance the translation of OA technology in clinical practice.

## Methods

### Linear model of OA signal acquisition

In an acoustically homogeneous medium, the propagation of the optoacoustically induced pressure wave $p(r, t)$ is described by the wave equation [4]:

$$\frac{\partial^2 p(r, t)}{\partial t^2} - c^2 \nabla^2 p(\boldsymbol{r}, t) = \Gamma \frac{\partial H(\mathrm{r}, \mathrm{t})}{\partial t}, \tag{1}$$

where $c$ is the speed of sound in the medium, $\Gamma$ is the dimensionless Grüneisen parameter, which is assumed constant, and $H$ is the energy absorbed in tissue per unit volume and per unit time, which is the source for OA fields. For pulsed laser illumination (thermal and stress confinement conditions [15]), the temporal dependence of $H$ can in the acoustic regime be approximated $\delta$-distribution, so that $H(r, t) = H_r(t)\delta(t)$.

The solution of Eq. 1 can be described in terms of Green's function for the wave equation [16, 17]. Under the assumption that an object $\Omega$ is an OA source, the solution of the wave equation for object $\Omega$ can be estimated by assuming that each point $\boldsymbol{r}'$ in the object is the emitter of OA waves, given by the Green's function $G$, which is given by:

$$G(\mathrm{r}, \mathrm{r}', \mathrm{t}) = \frac{\delta\left(t - \frac{|\boldsymbol{r} - \boldsymbol{r}'|}{c}\right)}{4\pi |\boldsymbol{r} - \boldsymbol{r}'|}. \tag{2}$$

Eq. 2 represents the elementary wave generated at $\boldsymbol{r}'$ at $t$=0, propagating outwards in a spherical shell. The solution to Eq. 1 at a given point $\boldsymbol{r}$ can be written as:

$$p(\mathrm{r}, \mathrm{t}) = \frac{\Gamma}{4\pi c^2} \frac{\partial}{\partial t} \left[ \int_{\Omega} H_r(\boldsymbol{r}') \frac{\delta\left(t - \frac{|\boldsymbol{r} - \boldsymbol{r}'|}{c}\right)}{|\boldsymbol{r} - \boldsymbol{r}'|} d\boldsymbol{r}' \right], \tag{3}$$

where the integral is estimated along the whole object. Eq. 3 represents the OA statement of the Huygens Principle, since it describes the OA pressure field generated by an object $\Omega$ at point $\boldsymbol{r}$ as a superimposition of elementary waves generated at points $\mathbf{r'}$ on the source.

**Model-based optoacoustic reconstruction**

Several algorithms have been suggested for the inversion of Eq. (3), i.e., for the estimation of the initial pressure $p_0$ or $H_r$, given the OA signal $p(r,t)$ at detector located in r. Back-projection and model-based algorithms are widely used for OA image formation. While the former is commonly used for its simplicity, the latter has been suggested as more precise reconstruction algorithms for accurate quantitative functional and molecular imaging in OA [5, 21].

Standard model-based algorithms reconstruct OA images by minimizing the difference between the measured OA signal and the signal theoretically predicted by the forward model. Typically, the broadband OA signal is entirely projected into one single image which contains structural information ranging from millimeter to micrometer resolution.

The discretization of Eq. (3) leads to a matrix equation of the form of [4]:

$$p = Mx, \tag{4}$$

where $p$ is the broadband pressure signal recorded; $x$ is the unknown distribution of optical absorption; and $M$ is the linear operator that maps the unknow initial pressure to the recorded signal. The inversion of Eq. (6) is achieved by minimizing the squared error:

$$x^* = \arg\min_x \|p - Mx\|_2^2. \tag{5}$$

A common problem when using standard model-based inversion is the appearance of negative values due to the bipolarity of the OA signal, which have no physical meaning since optical absorption can only be positive or zero. Furthermore, the OA inversion problem may be ill-posed leading to the appearance of additional spurious negative values introduced during the inversion process as part of the minimization. For these reasons, non-negative constraints are included in the algorithm to guarantee physical integrity of the solution [22], i.e., positive optical absorption:

$$x^* = \arg\min_{x \geq 0} \|p - Mx\|_2^2. \tag{6}$$

In contrast to unconstrained inversion, Eq. (6) cannot be solved analytically, and only iterative methods are applicable. Furthermore, since the problem in Eq. (6) is ill-posed and leads to inversion uncertainty, it requires additional regularization to impose additional constraints.

Thereby, Eq. (6) is modified by adding a Tikhonov regularization term, i.e.

$$x^* = \arg\min_{x \geq 0} \|p - Mx\|_2^2 + \lambda \|Lx\|_2^2, \tag{7}$$

where $\lambda$ is the regularize parameters, $\|\cdot\|_2$ is the $L^2$-norm. The matrix L can be selected as the identity matrix, which gives preference to solutions with a small norm, and penalizes any content in the data that is not in agreement with the model, like noise. Regularization is required to achieve a stable and unique solution.

The regularized problem in Eq. (7) can be formulated to the same form as Eq. (5):

$$x^* = \arg\min_{x \geq 0} f(x) = \arg\min_{x \geq 0} \left\| \tilde{p} - \widetilde{M} x \right\|_2^2, \tag{8}$$

with

$$\tilde{p} = \begin{pmatrix} p \\ 0 \end{pmatrix}$$

and

$$\widetilde{M} = \begin{pmatrix} M \\ \lambda L \end{pmatrix}$$

In this work, the Projected Conjugate Gradient [22] is used as non-negative least square (NNLS) methods to solve Eq. (8).

**Multiscale model-based reconstruction (fbMB)**

The new proposed multiscale model-based reconstruction (fbMB) framework allows to decompose the OA signal in different components and reconstruct multiple non-negative images with scale specific contrast. Soft priors on the frequency content of signals generated on different scales steer the acoustic spectrum towards higher or lower frequencies, while allowing overlaps when required by the system model. Thus, for multiscale reconstruction, the matrix equation in Eq. (4) is modified as follows:

$$p = M(x_1 + \cdots + x_n), \tag{9}$$

where $p$ is the broadband pressure signal recorded; and $x_1^*, \ldots, x_n^*$ are the unknown distributions of optical absorption at $n$ scales.

## Soft priors on frequency bands

In the framework of $L^2$ regularization, a priori information about the tissue being imaged can be included [23] to ensure convergence of the reconstruction algorithm to the correct optical absorption distribution on different scales. For this reason, soft priors on the frequency bands separation are added into the inversion. The priors are considered "soft" because they do not force a rigid frequency band separation, instead they allow overlap between frequency bands to preserve the physical accuracy of the solution.

The regularized inversion problem with soft priors can be written as:

$$(x_1^*, \ldots, x_n^*) \;=\; \underset{(x_1, \ldots, x_n) \geq 0}{\arg \min} \; \| p - M(x_1 + \cdots + x_n) \|_2^2$$

$$+ \lambda \| L(x_1 + x_2 + \ldots + x_n) \|_2^2$$

$$+ \eta \mu_1 \|(id - F_1)M x_1\|_2^2 + \cdots + \eta \mu_n \|(id - F_n)M x_n\|_2^2, \tag{10}$$

where $F_1, \ldots, F_n$ are the $n$ band-pass filters, and $\eta, \mu_1, \ldots, \mu_n$ the regularization parameters for the soft priors such that $\mu_1 + \cdots + \mu_n = 1$.

The regularized problem in Eq. (10) can be formulated in the same form as Eq. (8):

$$(x_1^*, \ldots, x_n^*) = \underset{(x_1, \ldots, x_n) \geq 0}{\arg \min} f(x_1, \ldots, x_n) = \underset{x \geq 0}{\arg \min} \left\| \tilde{p} - \tilde{M}(x_1 + \cdots + x_n) \right\|_2^2 \tag{11}$$

with

$$\tilde{p} = \begin{pmatrix} p \\ 0 \\ \vdots \\ 0 \end{pmatrix}$$

and

$$\widetilde{M} = \begin{pmatrix} M & \dots & \dots & M \\ \eta\mu_1(id - F_1)M & 0 & \dots & 0 \\ \vdots & \ddots & \ddots & \vdots \\ 0 & \dots & 0 & \eta\mu_n(id - F_n)M \end{pmatrix}.$$

The Projected Conjugate Gradient [21] is used as non-negative least square (NNLS) methods to solve Eq. (11). Numerical inversions are usually computationally demanding and imposing the additional non-negative constraint increases the inversion time. However, the computational complexity of fbMB is increased only by a constant factor (linear in the number of components) compared to standard model-based reconstruction, since the model needs to be evaluated once for every component $x_j$ and the filters need to be applied.

**Filters selection for the soft priors**

Standard Butterworth band-pass filters are used as soft priors $F_1, \dots, F_n$ in (Eq.11). The low and high pass cut-off frequencies, and the relative bandwidth are chosen to avoid ringing artifacts [6] such that:

$$\frac{BW_1}{f_{c1}} = \frac{BW_2}{f_{c2}} = \frac{BW_3}{f_{c3}} = \dots$$

In case of two scales decomposition, band-pass filters with frequency bands 0-1.23 MHz and 1.23-15 MHz are chosen in the soft priors, with 175% relative bandwidth. For three scales decomposition, band-pass filters with frequency bands 0-500kHz, 500kHz-1.23MHz and 1.23MHz-13.3MHz are selected, with a relative bandwidth of 136%.

Notably, the fbMB method is flexible because it is robust with respect to the choice of the filters used in the soft priors.

**MSOT system setup**

OA measurements were conducted using a commercially available, real-time, multispectral optoacoustic tomographic (MSOT) small animal scanner (inVision 256-TF, iThera Medical, Munich, Germany). The system features 360° ring illumination and 270° acoustic detection. The acoustic detection includes 256 cylindrically focused transducers with 5MHz central frequency and -6dB bandwidth of 150% (from the electrical impulse response reported in Supplementary Fig. 1.b) arranged on a circle of 40 mm radius to surround the sample (Fig. 1.a).

### *Ex vivo* mouse kidney imaging

OA images of an excised kidney from a nude mouse (4 weeks old) were acquired to demonstrate the structures enhancement of the new proposed algorithm fbMB in an organ with well-defined features of varying scales. Ten minutes after the animal euthanasia the kidney was excised and embedded in a supporting turbid agar gel in a cylindrical mold (12 mm in diameter and 2.5 cm in height). The agar gel was made by mixing 1.6% w/m agar gel (Agar for microbiology, Fluka analytical) with 0.8% v/v Intralipid 20% (Sigma).

### *In vivo* data acquisition

The multiscale capabilities of the new algorithm were tested with an *in vivo* mouse dataset, including 533 OA images from different anatomical sections, from the brain to the lower abdomen.

2D OA images were continuously acquired at single wavelength (800 nm) for anatomy visualization, or at multiple wavelengths (690-900 nm) for spectral unmixing quantification.

For 3D whole body imaging, 2D images of consecutive anatomical sections were collected with 200 µm resolution between adjacent frames and then stacked together for volumetric imaging. Animal procedures were approved by the Government of Upper Bavaria. Nude mice (3-4 weeks old) were anesthetized and placed lying prone in the animal holder such that the transducer array faced the ventral side. After data acquisition, the mice were euthanized by cervical dislocation.

### Signal pre-processing

In case of standard model-based reconstruction, the acquired OA signals are filtered prior to reconstruction by applying a Butterworth bandpass filter in the frequency range between 0.1 and 13.3 MHz to reject noise beyond the sensitivity of the transducers.

In case of simple filtering techniques, i.e., Butterworth and Wavelet, the OA signals are divided into different frequency bands by applying a bank of band pass filters and each frequency band reconstructed separately with standard model-based algorithm. In case of simple Butterworth filtering technique, the band-pass filters are chosen in agreement with the one selected in the soft priors (see Section 4.2.4). For the simple Wavelet filtering technique, the Matlab functions swt and iswt are used for two and three scales decomposition. The Daubechies mother wavelet with two vanishing moments 'db2' was selected as already reported in [24]. The advantage of using Wavelet decomposition over standard filters is that it does not require the a priori knowledge of the frequency information contained in the signal and guarantees optimal temporal and frequency resolution for each band.

**Residuals and Structural Similarity Index (SSIM)**

The efficiency of fbMB to separate different OA components was quantified against standard filtering techniques (i.e., Butterworth or Wavelet bandpass filters) in terms of physical and structural accuracy. For physical accuracy we refer to the capability of the algorithm to reconstruct OA images at different scales with the sum of them explains the acquired OA signals. For structural accuracy we indicate the capability of the algorithm to compensate for potential ripples and artifacts generated by the frequency band separations.

Herein, we use the normalized reconstruction residuals as physical accuracy indicators. They are obtained by dividing the reconstruction residuals of both fbMB and standard filtering techniques for the ones obtained with standard model-based reconstruction.

Moreover, we use the structural similarity index (SSIM) [25] between the reference images (reconstructed with standard model based) and the images obtained with the two different methods, i.e., fbMB and standard filtering techniques (sum of all components) to evaluate the structural accuracy.

**Image entropy and Perception-base Image Quality Evaluator (PIQE).**

To quantify image quality improvement of fbMB against standard model-based reconstruction, two different metrics are proposed: image entropy, as objective indicator of the image information content; perception-based image quality evaluator (PIQE) to account for human visual system perception [26]. The two indices were estimated on the three-channels RGB images.

To adjust image contrast, a histogram-based thresholding method was applied to all OA images by removing 0.5 % highest and 0.02% lowest pixel values. Note that the process was applied only to improve image visualization, and all quantification metrics were estimated on the raw images.

# Data availability

The data that support the findings of this study are available from the corresponding author upon reasonable request.

## Code availability

The code that support the findings of this study is available from the corresponding author upon reasonable request.

## Acknowledgements

This project has received funding from the European Union's Horizon 2020 research and innovation programme under the Marie Skłodowska-Curie grant agreement No 721766 (FBI) and from the European Research Council (ERC) under the European Union's Horizon 2020 research and innovation programme under grant agreement No 694968 (PREMSOT).

The authors would like to thank Dr. Robert J. Wilson for advice and editing, P. Anzenhofer for technical support and mouse handling and Dr. C. Zakian and S. Morscher for advice on MSOT imaging.

## Author Contributions

A.L. and D.J. conceived the idea of frequency-band model-based reconstruction for optoacoustic imaging. D.J. implemented the reconstruction algorithm. A.L. designed and performed all the experiments. All authors contributed to the writing of the manuscript.

## Competing Interests statement

## Supplementary Information

### Scale dependent SNR and frequency entanglement in OA

The analytical solution of the optoacoustic wave equation (Eq. 1 in Method section) for a uniformly illuminated sphere and reported in [17] explains the most important features of optoacoustic signals. First, the amplitude of the OA waves is directly proportional to the size of the object so that small scale features have lower SNR compared to big scale features (scale-dependent SNR). Second, the emission spectrum of the object is broadband, with peak emission frequency and bandwidth inversely proportional to the source size. Thus, the frequency content of small absorbers is more shifted towards the high frequencies. Third, the emission spectra of objects of different sizes located in the same field of view overlap (frequency entanglement) due to the linearity of Eq.1.

Supplementary Fig. 1c,d show the scale dependent SNR and the spectra entanglement in a multiscale OA phantom consisting of small-scale features ranging from 300 μm to 1.1 mm resolution (High f) embedded in a bulky absorbing object of 1.6 cm diameter (Low f). The generated OA signals from the phantom were obtained with a 2D simulation (K-wave, [18]) with a 40 μm grid resolution and 40 MHz time sampling, which enabled to simulate the propagation of acoustic waves in the frequency range 0-12 MHz.

Notably, from the frequency spectra reported in Supplementary Fig. 1d, the large object (Low f) emits characteristically stronger signal than the small-scale features (High f), and its frequency content is predominantly shifted toward the lower frequencies. Furthermore, due to the intrinsic broadband characteristic and linearity of the OA signals, the spectra of the two objects overlap.

These general characteristics of the OA signals, i.e., scale-dependent SNR and frequency entanglement, pose two challenges in OA imaging: being able to resolve small absorbers from large absorbing structures and disentangling the different OA scales in the signal domain.

**Frequency dependent sensitivity and out-of-plane signal**

Beside the scale-dependent SNR and the frequency entanglement, an additional challenge in OA imaging is represented by the detected out-of-plane signals which lead to artefacts in the reconstructed images. As light propagates in a highly scattering medium (as biological tissue is), it is diffused in all directions [19, 20] and an amount of optical energy is absorbed from chromophores located at different image planes, generating so-called out-of-plane signals. In OA tomography, cylindrically focused detectors are usually employed since they have a wide reception angle in the image plane (xy) and a narrow one in the elevational direction (yz) to reject out-of-plane signals (Supplementary Fig. 1c, Full BW) [1]. However, due to the frequency dependent sensitivity field, at low frequencies the detectors are still sensitive to the out-of-plane signals. Supplementary Fig. 1a shows the sensitivity fields of a cylindrically focused transducer in different frequency bands, low frequencies (Low f, 0-1.3 MHz) and high frequencies (High f, 1.3-13 MHz). While at high frequencies above 1.3 MHz (High f) the transducer has a good sensitivity in the image plane and a good focusing on the elevation direction, at low frequencies below 1 MHz (Low f) the detector is less focused in the elevational direction, and out-of-plane signals are detected.

Notably, the detected out-of-plane signals when projected into the reconstructed image results in structural distortion. Supplementary Fig. 1e illustrates an example of structural distortion for an OA phantom containing three absorbing microspheres, one positioned in the image plane (yellow sphere in Supplementary Fig. 1e) and two positioned at 1.7 mm and 3.4 mm far from the image plane (blue spheres in Supplementary Fig. 1e). As clearly reported in the reconstructed images, the out of plane spheres appear blurred and out of focus compared with the in-plane sphere. Furthermore, as reported in Supplementary Fig. 1e the detected OA signals from out of plane structures is less broadband and contains only low frequency components.

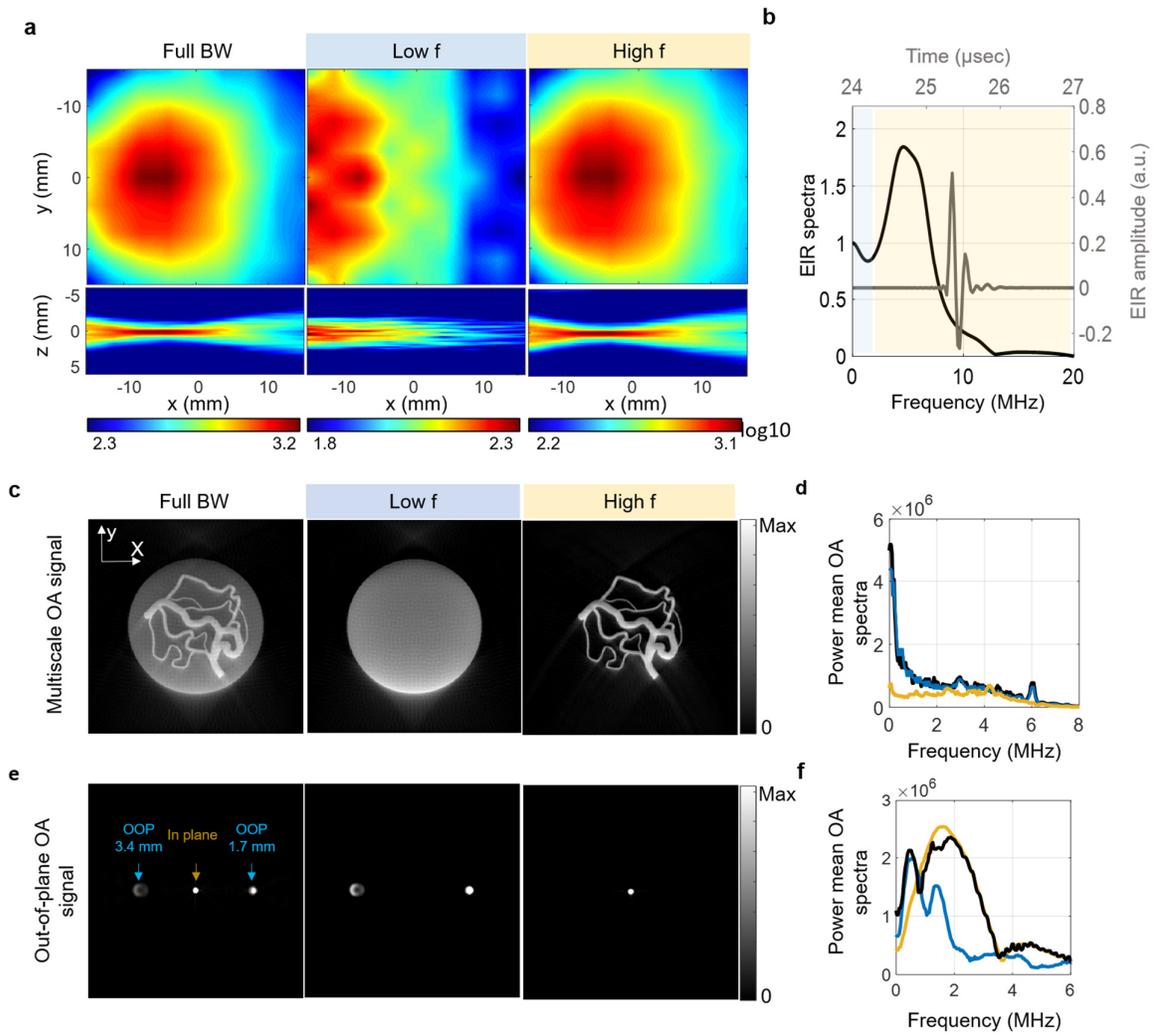

**Supplementary figure 1. Problem statement.** a) Frequency-dependent transducer sensitivity field in the image plane XY and in the elevational plane ZX. b) Transducer electrical impulse response (EIR) in time and frequency domain. c) OA image showing structures at different scales, i.e. bulky structure (Low f) and vessels-like structure (High f). d) Frequency overlap of different OA scales, bulky tissue in blue and vessels-like structure in yellow, and scale-dependent SNR with predominating low frequency energy. e) OA image showing in plane and 1.7 mm and 3.4 mm out-of-plane (OOP) microspheres. f) OA spectrum of the in-plane microsphere in yellow and low-frequency-shifted spectrum in blue for the out-of-plane microspheres. Full bandwidth (Full BW); Low frequencies (Low f); High frequencies (High f).